\documentclass{article}

\usepackage{arxiv}

\usepackage[utf8]{inputenc} 
\usepackage[T1]{fontenc}    
\usepackage{hyperref}       
\usepackage{url}            
\usepackage{booktabs}       
\usepackage{amsfonts}       
\usepackage{nicefrac}       
\usepackage{microtype}      
\usepackage{lipsum}

\usepackage{graphicx}

\title{Classical Music Prediction and Composition by means of Variational Autoencoders}

\author{
  Daniel Rivero \\
  Centro de investigación CITIC \\
Department of Computer Science\\
 University of A Coruña\\
 A Coruña, Spain \\
  \texttt{daniel.rivero@udc.es} \\
   \And
 Enrique Fernandez-Blanco \\
 Centro de investigación CITIC \\
 Department of Computer Science\\
 University of A Coruña\\
 A Coruña, Spain \\
  \texttt{efernandez@udc.es} \\
   \And
 Alejandro Pazos \\
 Department of Computer Science\\
 University of A Coruña\\
 A Coruña, Spain \\
  \texttt{apazos@udc.es} \\
}

\begin{document}
\maketitle

\begin{abstract}
This paper proposes a new model for music prediction based on Variational Autoencoders (VAEs). In this work, VAEs are used in a novel way in order to address two different problems: music representation into the latent space, and using this representation to make predictions of the future values of the musical piece. This approach was trained with different songs of a classical composer. As a result, the system can represent the music in the latent space, and make accurate predictions. Therefore, the system can be used to compose new music either from an existing piece or from a random starting point. An additional feature of this system is that a small dataset was used for training. However, results show that the system is able to return accurate representations and predictions in unseen data.
\end{abstract}



\section{Introduction}

Deep Learning has become an absolute revolution in art generation. Since the development of recent Deep Learning techniques, many advances have been published in this knowledge area. Most of them are focused on image generation 
\cite{2016arXiv160905473Y}, but other areas such as text generation have experienced much research \cite{2019arXiv190202192W}.

In music analysis, some works have begun to arise in recent years using Deep Learning techniques that allow the generation of audio. However, the most used methods (Generative Adversarial Networks, GANs, and Linear Shot-Term Memories, LSTMs), are very complex methods that require a very high training dataset with a long training time \cite{dong2018}\cite{Ycart2018}. Moreover, these models do not give any control to the user about the piece of music generated by the system. A system that allows the user to change the piece being generated at any time would be desirable.

Other models used for this task are Variational Autoencoders (VAEs) \cite{Higgins2017betaVAELB}. This model combines an encoder and a decoder in order to make a transformation from a usually high dimensional space into a lower dimensional space. This new space is called latent space, and its main feature is that any point from that latent space can be decoded returning an output with sense. Therefore, any latent vector between two (or more) musical pieces will return a musical piece with the properties of those mixed. This allows the user to change the values of the latent vector in order to get a similar musical piece. This feature can make VAEs a powerful technique for music composition, in which the user has more control and can change at any moment the melody being composed by the system.

VAEs have already been explored as systems for musical analysis \cite{esling2018}; however, its use for composition and prediction still has to be explored. This paper shows how VAEs can be used for music prediction and therefore for music composition from a small dataset. In this work, classical music was used for training the system and, as a result, the system is able to compose melodies with the same style as the dataset used for training. Moreover, and oppositely with other systems, VAEs allow the training with a relatively small dataset. We will show that a small number of musical pieces is enough for the system to learn a particular style of a composer. Results show that the system is able to perform in unseen data with the same accuracy as with data used in the training process. Therefore, no large datasets are needed to develop this system and to learn the style of an author.

The application of VAEs is done in a different way: usually, inputs and targets have the same value. In this case, different values have been used to allow the system to properly train to accomplish the objective of this work.

Also, even VAEs have already been used in the field of music analysis, the main approaches do not use pure VAEs with dense layers; instead of it, they are usually mixed with other techniques such as Convolutional Neural Networks (CNNs) \cite{yin_jyun_luo_2018_1492501}. This work presents an approach in which pure VAEs are used to perform classical music composition.

The rest of the paper is organised as follows: Section \ref{sec:seccionStateOfTheArt} contains a description of the most relevant works in this area. Section \ref{sec:seccionVAE} provides a description of VAEs. Section \ref{sec:seccionModel} describes the method used in this work, with the descripction of the representation of the music in \ref{sec:seccionRepresentation}, the implementation of the VAE in \ref{sec:AutoencoderImplementation}, and the different metrics used for measuring the results in \ref{sec:PerformanceMeasures}. Section 
\ref{sec:Experiments} describes the experiments carried out in this work. Finally, sections \ref{sec:seccionConclusions} and \ref{sec:seccionFutureWorks} describe the main conclusions of this work and the future works that can be carried out from it.


\section{State of the Art}

\label{sec:seccionStateOfTheArt}

Even music generation is an exciting task, up to date few works have focused on this task. These works use mainly Deep Learning techniques that allow new ways of signal processing. As new Deep Learning models have been developed, these models have been applied to music generation.

One of the most used techniques in art generation is Generative Adversarial Networks, which consist of two different neural networks: generator and discriminator. The generator is used to generate new plausible examples from the problem domain, while the discriminator is used to classify examples as real or fake. This model has been successfully applied to image generation \cite{goodfellow2014gan}. In the field of music generation, there are some works in which it has been applied, with some modifications to model the temporal structure and the multi-track interdependency of a song \cite{dong2017} \cite{yang2017} \cite{dong2018}.

Other works are focused on Long Short-Term Memory (LSTM) networks, which are networks with recurrent connections broadly used for temporal processing \cite{LSTM2001}. Simple LSTMs were used for musical transduction \cite{Ycart2018} to implement a pitch detection system. In another work, LSTMs were combined with VAEs for music generation \cite{roberts2018}. In this work, 2-layer LSTMs with 1024 neurons per layer were used for the decoder.

Autoencoders (AEs) have also been used as models in the field of music analysis. For instance, they were used to synthesize  musical notes, having as input raw audio instead of pitches\cite{2017arXiv170401279E}. Thus, the focus of this work was not to generate music compositions but independent raw musical notes.

VAEs have also been used for music-based tasks. For instance, in \cite{esling2018} and
\cite{2018arXiv180508501E}, VAEs were used for timbre studies. In this work, the Short-Term Fourier Transform (STFT) in combination with a Discrete Cosine Transform (DCT) and the Non-Stationary Gabor Transform (NSGT) to preprocess the audio before the application of the VAE. In another work, Convolutional Neural Networks have been used to extract features. However, its main goal was not to compose music, but to perform audio-to-MIDI alignment, audio-to-audio alignment, and singing voice separation \cite{yin_jyun_luo_2018_1492501}. 

In another work, VAEs are combined with LSTMs used as encoders, and recurrent neural networks (RNNs) as decoders \cite{roberts2018}. In a recent publication, VAES were used for sound modeling \cite{2018arXiv180604096R}. This work makes a comparison of different models, including VAEs, However, this work works with raw audio instead of MIDI files since its objective is to model sound and not music generation. VAEs have been used also for processing of speech signals with the objective of making modifications in some attributes of the speakers \cite{Blaauw2016} \cite{2017arXiv170404222H}


\section{Variational Autoencoders}

\label{sec:seccionVAE}

Variational Autoencoders arose from the evolution of autoencoders \cite{2013arXiv1312.6114K} \cite{Higgins2017betaVAELB} \cite{Kaae2016}. Both techniques aim to codify a set of data into a smaller vector, and then reconstructing the original data from this vector. In AEs, the set of data points in the input space is mapped to an smaller vector, which is a point in a new space with fewer dimensions called latent space. Usually, Multilayer Perceptrons (MLPs) are used for the codification and decodification tasks. This approach allows for interesting tasks such as information compression. However, in this latent space other points different than those obtained from an input do not usually correspond to an output with significance in the original space.

That is the reason why VAEs try to build a latent space, in which all of the latent vectors corresponding to inputs are close to each other, and the points in the space between them correspond to outputs that have a significance in the original space. VAEs allow powerful representations while being a simple and fast learning framework.

The objective of Variational Autoencoders is to find the underlying probability distribution of the data $p(x)$, where $x$ is a vector in a high dimensional space. In a lower-dimensional space $z$, a set of latent variables are considered. The model is defined by the probability distribution

\begin{equation}
\label{eq:eq1}
p(\mathbf{x},\mathbf{z}) = p(\mathbf{x}|\mathbf{z})p(\mathbf{z})
\end{equation}

The function $p(\mathbf{x}|\mathbf{z})$ represents a probabilistic decoder that models how the generation of observed data $\mathbf{x}$ is conditioned on the latent data $\mathbf{z}$. The function $p(\mathbf{z})$ represents the probability distribution of the latent space and is usually modeled by a standard Gaussian distribution.

The function $p(\mathbf{x}|\mathbf{z})$ can be approximated with a model $q(\mathbf{z}|\mathbf{x})$. This model works as an encoder, and, for a specific $\mathbf{x}$, emits two latent vectors, $\boldmath{\mu}$ and $\boldmath{\sigma}$, that represent the mean and standard deviation of the Gaussian probability for that data.

\begin{equation}
\label{eq:eq2}
q(\mathbf{z}|\mathbf{x}) = \mathcal{N}(\mathbf{z};\tilde{\mathbf{\mu}}(\mathbf{x}), \tilde{\mathbf{\sigma}}^2(\mathbf{x}))
\end{equation}

The optimization problem consists on minimizing the Kullback-Leibler (KL) divergence between the approximation and the original density. Using the Bayes' rule, the expression to minimize is the following:
\begin{equation} 
\label{eq:eq3}
{{D}_{KL}( q(\mathbf{z}|\mathbf{x}) || p(\mathbf{z}|\mathbf{x}))}  =   {\mathbb{E}_{q(\mathbf{z})} [log\,q(\mathbf{z}|\mathbf{x}) - log\,p(\mathbf{x}|\mathbf{z}) - log\,p(\mathbf{z}) + log\,p(\mathbf{x})]}
\end{equation} 
As it can be seen, the system is based on two parts: $q(\mathbf{z}|\mathbf{x})$ encodes the data into the latent representation, and  $\mathbf{z}$ $p(\mathbf{x}|\mathbf{z})$ is a decoder, generates data $\mathbf{x}$ from a latent vector $\mathbf{z}$. Since $p(z)$ does not depend on $q(z)$, this equation can be rewritten as the following

\begin{equation} 
\label{eq:eq4}
log\,p(\mathbf{x}) - {{D}_{KL}( q(\mathbf{z}|\mathbf{x}) || p(\mathbf{z}|\mathbf{x}))}  =   {\mathbb{E}_{q(\mathbf{z})} [log\,p(\mathbf{x}|\mathbf{z})]} - {{D}_{KL}( q(\mathbf{z}|\mathbf{x}) || p(\mathbf{z}))}
\end{equation} 

According to Eq.\ref{eq:eq4}, the objective can be changed into maximizing the marginal log-likelihood $log\,p_\theta(\mathbf{x})$ over a training dataset of vectors $\mathbf{x}$. This value to maximize is called the evidence lower bound (ELBO) and can be written as:

\begin{equation}
\label{eq:eq5}
log \ p(\mathbf{x}) = D_{KL}(q_(\mathbf{z}|\mathbf{x}) || p_\theta(\mathbf{z}|\mathbf{x})) + \mathcal{L}(\phi, \theta, \mathbf{x})
\end{equation}

In Eq.\ref{eq:eq5}, $\phi$ denotes the parameters of the encoder and $\theta$ the parameters of the decoder (weights and biases); $D_{KL}$ is the Kullback-Leibler divergence, non-negative, and $\mathcal{L}(\phi, \theta, \mathbf{x})$ is the variational lower bound. This value is calculated by the following equation

\begin{equation} 
\label{eq:eq6}
\mathcal{L}(\phi, \theta, \mathbf{x}) = \mathbb{E}_{q(\mathbf{z}|\mathbf{x})} [log\,p(\mathbf{x}|\mathbf{z})] - D_{KL}( q(\mathbf{z}|\mathbf{x}) || p(\mathbf{z}))
\end{equation} 

First term of Eq.\ref{eq:eq6} represents the average accuracy obtained by the system when using an approximate q instead of p. The second term represents the error made by using $q(\mathbf{z}|\mathbf{x})$ instead of $p(\mathbf{z})$ and allows to regularize the approximation q to be close to the true distribution.

Many times a $\beta$ value is introduced in the second term, leading to the $\beta$-VAE formulation. This modification can lead to having better results \cite{Higgins2017betaVAELB}; however, some works suggest a modification of this parameter through the training process \cite{Kaae2016}. This term allows the control of the trade-off between output signal quality and compactness/orthogonality of the latent coefficients $\mathbf{z}$.

The first term of the previous equation is approximated with the average value of the calculation of $log\,p(\mathbf{x}|\mathbf{z_l})$, where for each data $\mathbf{x}$ in the training set $\mathbf{z_l}$ are samples from the distribution $q(\mathbf{z}|\mathbf{x})$. Thus, the calculation of this term becomes
\begin{equation} 
\label{eq:vlb}
\frac{1}{L}\sum_{l=1}^L{log\,p(\mathbf{x}|\mathbf{z_l})}
\end{equation} 
The values $\mathbf{z_l}$ are taken from the distribution $\mathcal{N}(\mu(\mathbf{x}), \sigma(\mathbf{x}))$, where $\mu(\mathbf{x})$ and $\sigma(\mathbf{x})$ are the ouputs from the decoder with $\mathbf{x}$ as input.


\section{Model}

\label{sec:seccionModel}

\subsection{Representation}

\label{sec:seccionRepresentation}

The model proposed in this paper uses a representation of the music with the shape of a binary matrix $M$ with dimensions $n$x$t$, where $n$ is the number of pitches and $t$ is time. In this work, a time step of 100 milliseconds was used. Therefore, $M_{ij} = 1$ for those moments $j$ in which a pitch $i$ is being played, and 0 when that pitch is not being played. In this codification, the velocity (i.e., the volume) of the note events has not been taken into consideration.

This matrix is built from a MIDI file that contains the notes and the duration of each one. From these files, the notes are read and the matrix $M$ is built. A note with a pitch i that begins in the instant t and has a duration of $100*d$ milliseconds will be situated in row i, and in the columns from $t$ to $t+d$. In the case of having two consecutive overlapping notes with the same pitch, the matrix $M$ will have, on that note, a series of 1s from the beginning of the first note to the end of the second, with no distinction of being one long note or two different overlapping notes. To make this distinction, the value of $M$ at the moment previous to the second note is set to 0.

To make the dataset, 14 different compositions of Handel were used. Although it may seem a very low number, one of the objectives of this work is to show that this system can learn the representation into the latent space with a small dataset, and the resulting model can learn correctly the style of an author. These compositions were codified into binary matrices as described before.

\subsection{Autoencoder implementation}
\label{sec:AutoencoderImplementation}

Once the matrices were obtained, the inputs and targets of the VAE were elaborated. In a traditional approximation, the inputs and targets of the VAE are the same vectors. However, in this particularly case, a different approach was used. The inputs were a time window of $T$ seconds of the matrix $M$, i.e., a set of consecutive columns reshaped as a vector. An overlapping of $T-1$ seconds was used for building different inputs from the same matrix. In this case, the targets for these inputs were not the same vector. Instead of it, the target for each input  was the following input. Since an input and the following one had a difference of 1 second, $T-1$ seconds of the targets correspond to values in the inputs, and 1 second of the targets are new musical notes.

With this approach, we want the VAE not only to learn the dependencies between notes that allow making a representation in the latent space, but also the dependencies with the next second of music. Therefore, VAE is aimed to solve two different problems: music recomposition and music prediction. Once $T$ seconds of music are codified into a latent vector, the decodification of this vector returns the following $T$ notes with overlapping of $T-1$ seconds. Therefore, with this VAE trained, the generation of new musical compositions is very simple. First, $T$ initial seconds of music are taken. These seconds may be some already existing seconds from any composition or the decodification of any random vector in the latent space. Then, these seconds are used as inputs to the VAE to generate $T$ seconds in which the first $T-1$ are overlapping. Thus, 1 second of new music is generated. These $T$ seconds can be used again as inputs to generate a new second, and therefore a loop is built in which it can have as many iterations as seconds are needed.

It is important to bear in mind that this system does not compose songs with a structure. Instead of it, it is able to complete a composition in those parts in which the music may be missing. If the system is used for this purpose, then the $T$ initial seconds will be an existing part of the composition, and the system will try to continue this composition in the same way as Handel would have done.

Moreover, the part given by the system is not unique. Different parts can be returned if the vector is modified in the latent space. In this sense, this system allows the composition of different melodies.

Since a VAE is used, a loss function has to be given. This loss function was defined in section \ref{sec:seccionVAE}. In this implementation, sigmoid functions are used in the output layer, and therefore the system returns values between 0 and 1. In order to build music representation as a boolean matrix as described in section \ref{sec:seccionRepresentation}, a threshold must be used. Experimentation with different thresholds has to be done in order to find one the returns the best results.

\subsection{Performance measures}
\label{sec:PerformanceMeasures}

In order to measure the behaviour of this system, the outputs generated by this system, after the application of the threshold, must be compared with the targets. Since both outputs and targets are boolean values, this comparison can be done by means of Accuracy ($ACC$), Sensitivity ($SEN$) and Predictive Positive Value ($VPP$).

\begin{figure}
    \centerline{\includegraphics[width=12cm]{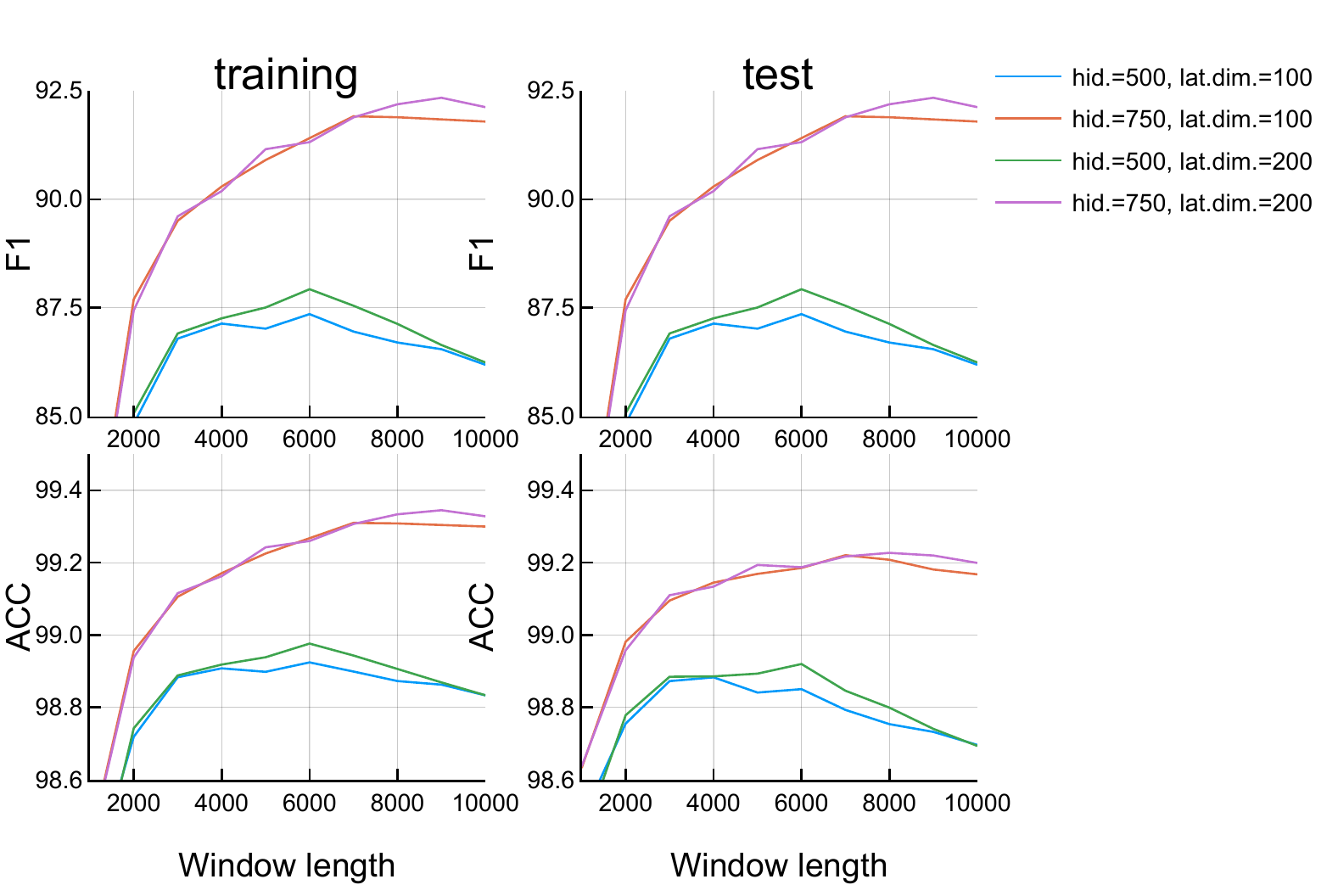}}
    \caption{F1 score and accuracy for each configuration}
    \label{fig:1}
\end{figure}

These values can be calculated from a confusion matrix. This matrix is built from 4 values:

\begin{itemize}
\item True Positives ($TP$) is the number of pitches and time steps correctly played.
\item False Positives ($FP$) is the number of pitches played in a time step in which there should be silence.
\item True Negatives (TN) is the number of pitches and time steps in which no notes are played and there should be silence.
\item False Negatives ($FN$) is the number of pitches and time steps in which no notes are played but there should be played.
\end{itemize}

From these values, $ACC$, $SEN$ and $VPP$ can be calculated with the following equations \cite{FAWCETT2006861}:

\begin{equation}
\ ACC = {\frac {TP+FP}{TP+FP+TN+FN}}
\end{equation}

\begin{equation}
\ SEN = {\frac {TP}{TP+FN}}
\end{equation}

\begin{equation}
\ VPP = {\frac {TP}{TP+FP}}
\end{equation}

Accuracy ($ACC$) represents the ability of the model to play values adequately. Sensitivity ($SEN$) represents the ability to play the true notes, even some additional notes might be played. Predictive Positive Value ($PPV$) is the probability of the system that the notes played correspond to true notes, even if some true notes are left to be played.

Since $SEN$ and $PPV$ are important measures, a good trade-off between $SEN$ and $PPV$ is needed. These values, as well as $ACC$, highly depend on the threshold chosen. A low value on the threshold leads to having a high number of notes played, even if many of them do not correspond to the original piece. A high value of the threshold corresponds to having a low number of notes, but with a high probability that they belong to the original piece.

Many times, these two measures ($SEN$,$PPV$) are summarized in a single metric called $F1-score$. This metric is with the harmonic mean of $SEN$ and $PPV$ metrics and it is usually better than accuracy on imbalanced binary data, as is the case of this dataset \cite{hastie_09_elements-of.statistical-learning}.


\section{Experiments}
\label{sec:Experiments}

In order to develop the system described in section \ref{sec:AutoencoderImplementation}, different experiments were carried out to set the values of the different parameters.

As it was already said, the system tries to predict 1 second of music from $T-1$ seconds of music. Therefore, this value of $T$ is an important parameter. Low values doe not give enough information to predict the music. On the other hand, too large values may give too much information and make the training process too slow, and overfit to the training set. The experimentation was performed with values of $T=2$ to $T=10$ seconds. Additionally, a value of $T=1$ was also chosen, in order to make predictions of 0.5 seconds instead of 1 second.

With respect to the architecture of the network, in all of the cases, the decoder was a MLP with one hidden layer and the decoder had the same architecture. An important parameter is the number of neurons in the hidden layers. Experiments with 500 and 750 neurons were performed.

Another important parameter was the dimension of the latent space. In this sense, values of 100 and 200 were chosen for this parameter. Finally, a value of $\beta = 0.5$ was used in these experiments.

The dataset was divided leaving the 20\% of the data for the test. This 20\% for the test was chosen to be in compositions different from the training set.

The system was trained with all of the parameter configurations described. For each configuration, different thresholds were used in order to select the best threshold and configurations. Figure \ref{fig:1} shows the results obtained for each configuration. The figures on the left show the $F1$-scores and accuracies obtained with the best threshold for each configuration on the training set. The figures on the right show the $F1$-scores and accuracies obtained on the test set with the thresholds given by the left figures. As it can be seen, accuracies and $F1$-scores are very high on both training and test sets. From this figure, a configuration with a window size of 9 seg, 750 hidden neurons and a latent dimension of 200 was chosen since it returned the best results on the training dataset.

\begin{figure}
    \centerline{\includegraphics[width=12cm, height=6cm]{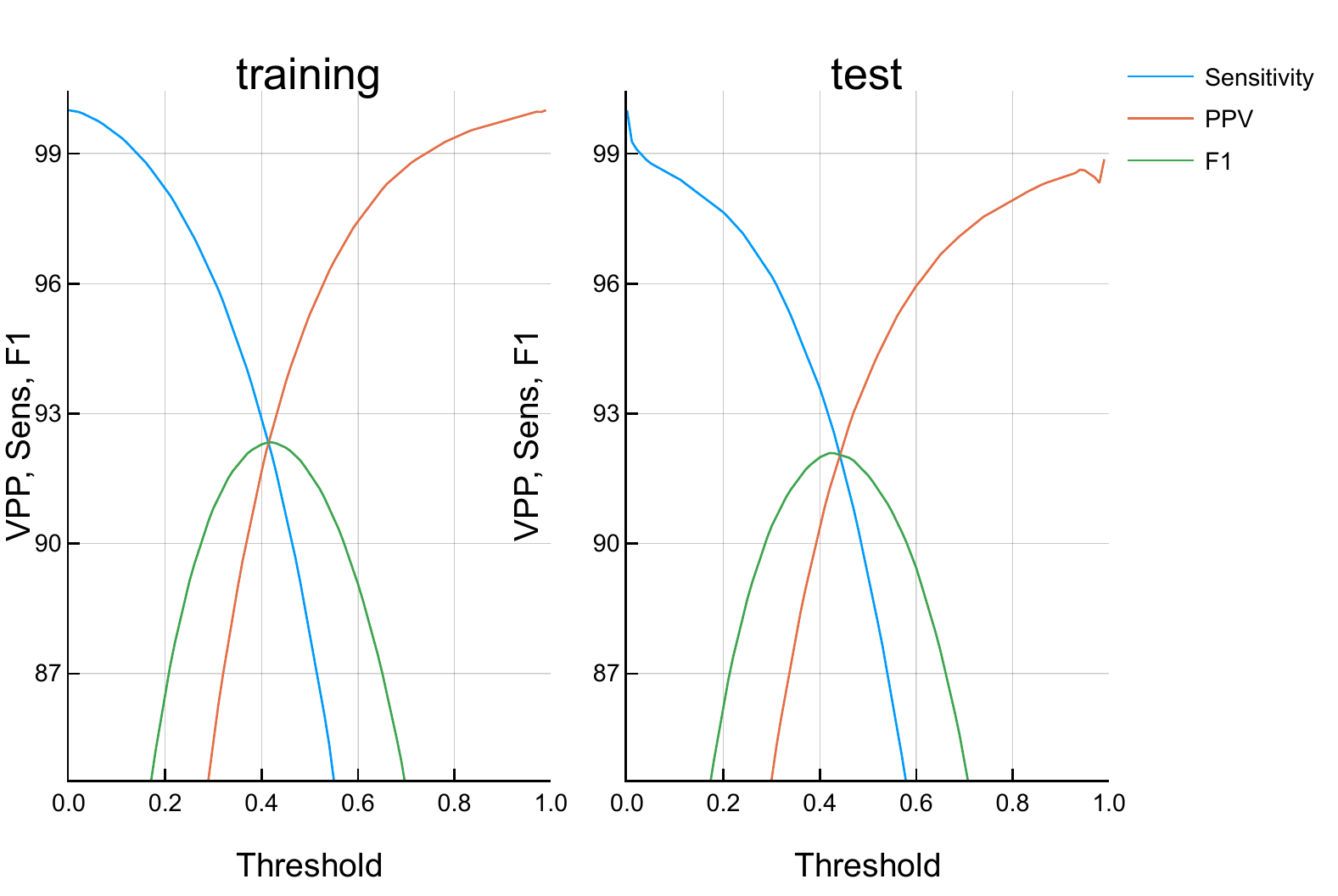}}
    \caption{Sensitivity, $VPP$ and $F1$-score for the different threshold in the selected configuration}
    \label{fig:2}
\end{figure}

Figure \ref{fig:2} shows the results obtained with a different threshold for this configuration. This figure shows Sensitivity, $PPV$ and $F1$-score on training and test sets. As  can be seen, low threshold values lead to having high sensitivity and low $PPV$. On the other side, a high threshold leads to having low sensitivity and high $PPV$. Therefore, a trade-off between these two values is needed. For this reason, $F1$-score was used. As a result, the value with the highest $F1-score$ in the training set was used as a threshold, being this value 0.41. The right plot shows the sensitivity, $PPV$ and $F1$-score for the test set.

As these figures show, the test results are very close to the training results, even when having a small dataset, and very low overfitting is observed in these graphs. This leads to the conclusion that the system has learned the features from a small set of compositions from this author, and the representation of this pieces is robust and can be applied to new compositions from the same author.

\begin{figure}
    \centerline{\includegraphics[width=12cm]{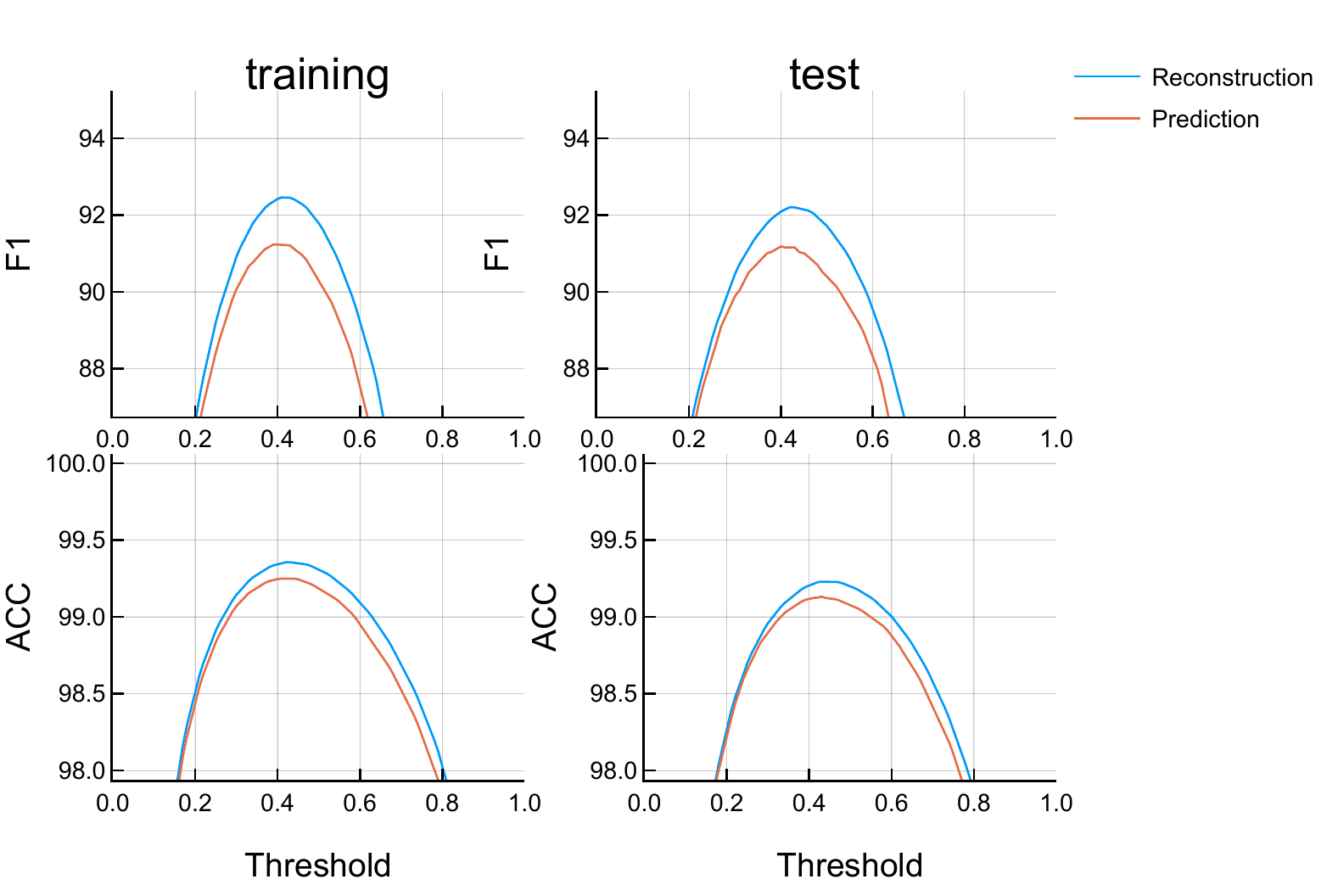}}
    \caption{Comparison of the results in the reconstruction and prediction problems}
    \label{fig:3}
\end{figure}

However, in this work, the application of VAEs is performed not only to codify and reconstruct a part of a composition, but also to predict the next second of the composition. Therefore, it can be seen that two different problems are studied here: reconstruction and prediction. Figure \ref{fig:3} shows the $F1$ scores and accuracies obtained for the selected configuration and threshold value, measured separately for the 8 seconds reconstruction and the 1-second prediction. As was expected, the $F1$ score and accuracy of the prediction are lower than those of the reconstruction. However, it is slightly lower, with a small difference. This figure shows training results on the left and test results on the right. Therefore, the system shows a better behaviour when making a prediction of a musical part from this author.

\begin{figure}
    \centerline{\includegraphics[width=12cm]{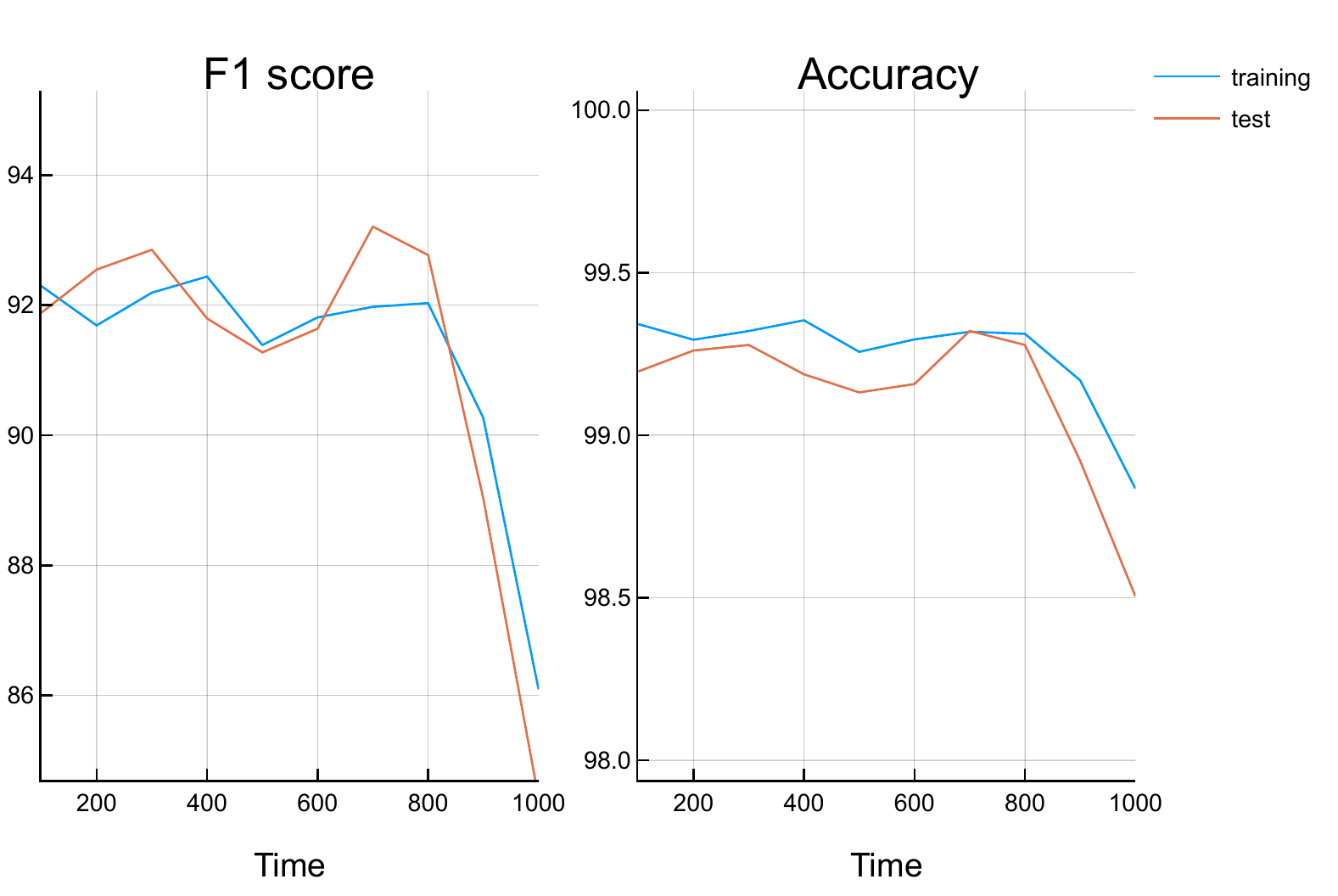}}
    \caption{Results in prediction on different moments}
    \label{fig:4}
\end{figure}

According to the results shown in Figure \ref{fig:3}, the prediction of music is very accurate to the real music. In this sense, it is interesting to see if this prediction is better or worse as the prediction time is further from the last known moment. Figure \ref{fig:4} shows the $F1$ score and accuracies obtained for the different moments in the one-second prediction. As a timestep of 100 milliseconds was used, this figure only has 10 values n the x-axis. As it can be seen in this figure, $F1$ score and accuracy seem not to be influenced by the moment of the prediction until it reaches the one second limit. At this moment, the prediction drops.


\section{Conclusions}
\label{sec:seccionConclusions}

This work shows the possibility of using VAEs for music analysis to solve two different problems: music codification and reconstruction, and music prediction. Two different metrics have been used, and, as the results show, both problems have been successfully solved since the results are very high in both metrics. As could be imagined, the prediction problem shows worse results than the reconstruction problem. However, these results are only slightly worse. This means that the learned features can make accurate predictions.

Therefore, it can be concluded that the features learned from the VAE correctly represent the style of the author. Moreover, this system was developed so that the decodification of the features corresponding to a specific piece leads to having the following piece of music. As this next piece of music can be codified and decodified into the following and so on, a composition can be represented in the latent space as a trajectory between different vectors in this space. Further analysis of these trajectories can give new insights about the composition and style of different authors.

The development with a small dataset is one of the most prominent features of this work. Although this could be considered as a big drawback for the training of the VAE, test results are comparable to training results. Therefore, the systems behave correctly with unseen pieces of music, returning their features in the latent space, and giving a prediction of the following second.

\section{Future Works}
\label{sec:seccionFutureWorks}

From the work presented in this paper, different directions can be taken. First, in the modeling of the audio, the velocity (volume) has not been taken into account. A new model could be developed in which the output of each neuron, a real value between 0 and 1, can be interpreted as the volume of that pith in that specific moment.

As was already explained in  section \ref{sec:seccionConclusions}, a musical composition can be represented as a trajectory in the latent space. This system could be trained with compositions from different authors. When high enough accuracies were obtained in both training and test sets, these trajectories could be analyzed in order to discover the differences between authors. This could be mixed with a clustering technique to discover interesting patterns in music composition.

Finally, as shown in section \ref{sec:Experiments}, the prediction seems to drop when it is getting to 1 second. Further studies should be carried out in order to find out if this prediction can be improved with bigger network architecture.

\bibliographystyle{unsrt}
\bibliography{references} 

\end{document}